\documentclass[a4paper,aps,twocolumn,nofootinbib]{revtex4}
\RequirePackage[colorlinks,hyperindex]{hyperref}
\RequirePackage[english]{babel}
\RequirePackage[latin1]{inputenc}
\RequirePackage[T1]{fontenc}
\RequirePackage{mathrsfs}
\RequirePackage{amsmath}
\RequirePackage{amssymb}
\RequirePackage{amsbsy}
\RequirePackage{color}
\RequirePackage{bm}
\hypersetup{colorlinks=true,breaklinks=true,urlcolor=blue,linkcolor=red}
\begin{document}
\title{\bf{Non-Trivial Effects of Sourceless Forces for Spinors:\\ Toward an Aharonov-Bohm gravitational effect?}}
\author{Luca Fabbri$^{1}$, Flora Moulin$^{2}$, Aur\'{e}lien Barrau$^{2}$}
\affiliation{$^{1}$DIME Sezione di Metodi e Modelli Matematici, 
Universit\`{a} degli studi di Genova,\\
via all'Opera Pia 15, 16145 Genova, ITALY\\
$^{2}$Laboratoire de Physique Subatomique et de Cosmologie, 
Universit\'{e} Grenoble-Alpes CNRS/IN2P3,\\
53 avenue des Martyrs, 38026 Grenoble cedex, FRANCE}
\date{\today}
\begin{abstract}
Spinor fields are written in polar form so as to compute their tensorial connection, an object that contains the same information of the connection but which is also proven to be a real tensor. From this, one can still compute the Riemann curvature, encoding the information about gravity. But even in absence of gravity, when the Riemann curvature vanishes, it may still be possible that the tensorial connection remains different from zero, and this can have effects on matter. This is shown with examples in the two known integrable cases: the hydrogen atom and the harmonic oscillator. The fact that a spinor can feel effects due to sourceless actions is already known in electrodynamics as the Aharonov-Bohm phenomenon. A parallel between the electrodynamics case and the situation encountered here will be drawn. Some ideas about relativistic effects and their role for general treatments of quantum field theories are also underlined.
\end{abstract}
\maketitle
\section{Introduction}
Quantum field theory (QFT) is one of the most impressive successes of contemporary science. From the standard model of particle physics to condensed matter theory, this framework works remarkably well and delivers high-precision predictions. The mathematical foundations of QFT however remain quite confusing. Some of the best known problems are the following (see \cite{Fabbri2019}): all calculations are performed by expanding fields in plane waves, which are not square integrable (and do not really exist as physical objects); in this expansion the coefficients are interpreted as creation and annihilation operators, lacking a precise definition \cite{streater}; and the calculations rely on the so-called interaction picture, which is in tension with the concept of a Lorentz-covariant field theory \cite{haag}. For all those reasons, it is clearly meaningful to consider a more general framework than ordinary QFT. This is the setting used in this work. As QFT works extremely well in all known situations, possible new results will obviously arise only in subtle cases.

As it is well known, Dirac spinor fields can be classified using the so-called Lounesto classification according to two classes: singular spinor fields are those subject to the conditions $i\overline{\psi}\boldsymbol{\pi}\psi\!=\!0$ and $\overline{\psi}\psi\!=\!0$ while regular spinor fields are all those for which the two above conditions do not identically hold \cite{L, Cavalcanti:2014wia, HoffdaSilva:2017waf, daSilva:2012wp, daRocha:2008we, Villalobos:2015xca, Ablamowicz:2014rpa}. For the regular spinor fields, it is possible to perform what is known as the polar decomposition of the Dirac spinor field \cite{Fabbri:2016msm}: this is the way in which it can be written in the Madelung form, that is with all the complex quantities expressed as a real module times a unitary complex exponential (\ref{spinor}) while respecting the transformation properties of a $1/2$-spin spinor field. In this form, the $8$ real components of spinors are re-arranged so as to show the physical information: of these $8$ components in fact, $3$ are shown to be the spatial directions of the velocity, $3$ are the spatial directions of the spin, $1$ is the usual expression of the module, and a last $1$ is a phase shift between left-handed and right-handed chiral parts of the spinor. This exhibits a possible internal dynamics, not taken into account in QFT. New effects can be associated with this phase.

Details about the spinor field equations in this form can be found in \cite{Fabbri:2016laz}. By implementing the Madelung form, so as to write every spinorial component as a module times a unitary exponential, and using the Gordon decompositions, so as to respect covariance, it is possible to convert the Dirac spinor field equation into a pair of coupled and non-linear vector field equations which are equivalent to the Dirac one.

These field equations determine the dynamics and the structure of the degrees of freedom of the spinor field in terms of two quantities collectively called the tensorial connection since they are built in terms of the connection but are also proven to be real tensors \cite{Fabbri:2017pwp}. In \cite{Fabbri:2018crr}, we eventually proved that with the tensorial connection it is possible to calculate the Riemann tensor, which represents the space-time curvature thus deciphering the information about the gravitational field.

In absence of gravitation the space-time curvature vanishes, and the Riemann tensor becomes zero identically. In this case, just as the connection, the tensorial connection may still be different from zero, but just like any tensor, if the tensorial connection happens to be non-zero then it will remain such in any system of reference: if this were to happen, we would be in presence of an object which, on the one hand, would represent a potential having a non-trivial structure, while on the other hand, it would have a vanishing strength.

This circumstance is the sourceless case, that is when the gravitational impact of the considered matter is identically zero (the Riemann tensor vanishes, so the Ricci tensor vanishes, which means that the energy density is not large enough to source gravity). Nevertheless, an influence on matter can still arise if the tensorial connection is not identically equal to zero.

As far-fetched as this situation may look, we will show that it is indeed what happens in two notable examples, given by the two integrable cases we know: the hydrogen atom and the harmonic oscillator.

These two examples, both from some remarkable physical potentials, and both exact solutions, should convince the skeptical reader of the fact that the structure of the wave function of a relativistic quantum matter distribution is in fact due to the non-vanishing tensorial connection even when it has no space-time curvature.

One should also keep in mind that a similar situation is already known. In the same way in which a relativistic quantum matter distribution can be affected by a non-vanishing connection, even when it has no space-time curvature, it can also be affected by some non-zero potential, even when it has no gauge curvature. This is the Aharonov-Bohm effect, which happens when wave functions display a phase-shift due to potentials even in regions where they give rise to no electrodynamic forces. Thus, in a way, we may say that what we are going to present consists in exhibiting the effects on matter of a gravitational Aharonov-Bohm effect.

This effect for gravity seems to be richer than for electrodynamics as in this case the full wave function, and not only its phase, can be modified. A comparative analysis of the two Aharonov-Bohm effects will be given.

As a bonus, we will show how it could be possible to obtain, in analogy to the Born rule for the discretization of electrodynamic degrees of freedom, a kind of Born rule for the discretization of gravitational degrees of freedom.

Some comments regarding the non-relativistic limit will eventually be sketched in one final section.
\section{Polar Spinors}
\subsection{Kinematic Quantities}
We will consider the Clifford matrices $\boldsymbol{\gamma}^{a}$ from which $\left[\boldsymbol{\gamma}_{a}\!,\!\boldsymbol{\gamma}_{b}\right]\!=\!4\boldsymbol{\sigma}_{ab}$ and $2i\boldsymbol{\sigma}_{ab}\!=\!\varepsilon_{abcd}\boldsymbol{\pi}\boldsymbol{\sigma}^{cd}$ defining the $\boldsymbol{\sigma}_{ab}$ and $\boldsymbol{\pi}$ matrices (this latter is what is usually called $\boldsymbol{\gamma}^5$ or $\boldsymbol{\gamma}_5$ with a sign ambiguity that has to be fixed by convention).

As known, Clifford matrices account for a total of $16$ linearly independent generators for the space of $4 \times 4$ complex matrices, given by
\begin{eqnarray}
\mathbb{I}, \ \ \boldsymbol{\gamma}^{a}, \ \ \boldsymbol{\sigma}^{ab}, \ \ \boldsymbol{\pi}, \ \ \boldsymbol{\gamma}^a\boldsymbol{\pi}
\label{base}
\end{eqnarray}
and it is possible to prove that they verify
\begin{eqnarray}
&\boldsymbol{\gamma}_{i}\boldsymbol{\gamma}_{j}\boldsymbol{\gamma}_{k}
=\boldsymbol{\gamma}_{i}\eta_{jk}-\boldsymbol{\gamma}_{j}\eta_{ik}+\boldsymbol{\gamma}_{k}\eta_{ij}
+i\varepsilon_{ijkq}\boldsymbol{\pi}\boldsymbol{\gamma}^{q}
\end{eqnarray}
which is a spinorial matrix identity (notice that this identity shows the pseudo-scalar character of the $\boldsymbol{\pi}$ matrix).

Given the spinor field $\psi$, its complex conjugate spinor field $\overline{\psi}$ is defined in such a way that bi-linear quantities
\begin{eqnarray}
&\Sigma^{ab}\!=\!2\overline{\psi}\boldsymbol{\sigma}^{ab}\boldsymbol{\pi}\psi\\
&M^{ab}\!=\!2i\overline{\psi}\boldsymbol{\sigma}^{ab}\psi
\end{eqnarray}
with
\begin{eqnarray}
&S^{a}\!=\!\overline{\psi}\boldsymbol{\gamma}^{a}\boldsymbol{\pi}\psi\\
&U^{a}\!=\!\overline{\psi}\boldsymbol{\gamma}^{a}\psi
\end{eqnarray}
as well as
\begin{eqnarray}
&\Theta\!=\!i\overline{\psi}\boldsymbol{\pi}\psi\\
&\Phi\!=\!\overline{\psi}\psi
\end{eqnarray}
are all real tensors, and it is possible to prove that they verify
\begin{eqnarray}
&\Sigma^{ab}\!=\!-\frac{1}{2}\varepsilon^{abij}M_{ij}\\
&M^{ab}\!=\!\frac{1}{2}\varepsilon^{abij}\Sigma_{ij}
\end{eqnarray}
together with
\begin{eqnarray}
&M_{ab}\Phi\!-\!\Sigma_{ab}\Theta\!=\!U^{j}S^{k}\varepsilon_{jkab}\label{A1}\\
&M_{ab}\Theta\!+\!\Sigma_{ab}\Phi\!=\!U_{[a}S_{b]}\label{A2}
\end{eqnarray}
alongside to
\begin{eqnarray}
&M_{ik}U^{i}\!=\!\Theta S_{k}\label{P1}\\
&\Sigma_{ik}U^{i}\!=\!\Phi S_{k}\label{L1}\\
&M_{ik}S^{i}\!=\!\Theta U_{k}\label{P2}\\
&\Sigma_{ik}S^{i}\!=\!\Phi U_{k}\label{L2}
\end{eqnarray}
and also
\begin{eqnarray}
&\frac{1}{2}M_{ab}M^{ab}\!=\!-\frac{1}{2}\Sigma_{ab}\Sigma^{ab}\!=\!\Phi^{2}\!-\!\Theta^{2}
\label{norm2}\\
&\frac{1}{2}M_{ab}\Sigma^{ab}\!=\!-2\Theta\Phi
\label{orthogonal2}
\end{eqnarray}
and
\begin{eqnarray}
&U_{a}U^{a}\!=\!-S_{a}S^{a}\!=\!\Theta^{2}\!+\!\Phi^{2}\label{norm1}\\
&U_{a}S^{a}\!=\!0\label{orthogonal1}
\end{eqnarray}
called Fierz re-arrangement identities.

These identities are important because in the general case of regular spinors, for which $i\overline{\psi}\boldsymbol{\pi}\psi\!\neq\!0$ or $\overline{\psi}\psi\!\neq\!0$, we can use (\ref{norm1}) to see that the $U^{a}$ vector is time-like. Three boosts can therefore be used to remove its spatial components and two rotations can be used to rotate $S^{a}$ along the third axis, while the third one eliminates the general phase. When these operations are performed, the most general spinor field compatible with those restrictions is
\begin{eqnarray}
&\!\psi\!=\!\phi e^{-\frac{i}{2}\beta\boldsymbol{\pi}}
\boldsymbol{S}\left(\!\begin{tabular}{c}
$1$\\
$0$\\
$1$\\
$0$
\end{tabular}\!\right)
\label{spinor}
\end{eqnarray}
in chiral representation. The matrix $\boldsymbol{S}$ is a generic complex Lorentz transformation, the angle $\beta$ is called Yvon-Takabayashi angle and represents the phase shift between right-handed and left-handed chiral parts of the spinor while $\phi$ is the module. The full spinor field is then said to be in \emph{polar form} \cite{Fabbri:2016msm}. In this polar form, the two antisymmetric tensors reduce to 
\begin{eqnarray}
&\Sigma^{ab}\!=\!2\phi^{2}(\cos{\beta}u^{[a}s^{b]}\!-\!\sin{\beta}u_{j}s_{k}\varepsilon^{jkab})\\
&M^{ab}\!=\!2\phi^{2}(\cos{\beta}u_{j}s_{k}\varepsilon^{jkab}\!+\!\sin{\beta}u^{[a}s^{b]})
\end{eqnarray}
with the two vectors
\begin{eqnarray}
&S^{a}\!=\!2\phi^{2}s^{a}\\
&U^{a}\!=\!2\phi^{2}u^{a}
\end{eqnarray}
and the two scalars
\begin{eqnarray}
&\Theta\!=\!2\phi^{2}\sin{\beta}\\
&\Phi\!=\!2\phi^{2}\cos{\beta}
\end{eqnarray}
in terms of the Yvon-Takabayashi angle and module.

All Fierz identities trivialize except for
\begin{eqnarray}
&u_{a}u^{a}\!=\!-s_{a}s^{a}\!=\!1\\
&u_{a}s^{a}\!=\!0
\end{eqnarray}
which show that the velocity and the spin are constrained, so that in general they amount to $3$ components each. The most general spinor therefore possesses $4$ components, or $8$ real components, given by the $3$ real components of the velocity and the $3$ real components of the spin, which can always be boosted or rotated away, plus the Yvon-Takabayashi angle and module, whose scalar character makes them impossible to be removed with a choice of frame. The latter are therefore the only $2$ real degrees of freedom of the spinor field.

From the metric, we define the symmetric connection as usual with $\Lambda^{\sigma}_{\alpha\nu}$ from which, with the tetrads, we define the spin connection $\Omega^{a}_{b\pi}\!=\!\xi^{\nu}_{b}\xi^{a}_{\sigma}(\Lambda^{\sigma}_{\nu\pi}\!-\!\xi^{\sigma}_{i}\partial_{\pi}\xi_{\nu}^{i})$. With the gauge potential, we then define the spinor connection
\begin{eqnarray}
&\boldsymbol{\Omega}_{\mu}
=\frac{1}{2}\Omega^{ab}_{\phantom{ab}\mu}\boldsymbol{\sigma}_{ab}
\!+\!iqA_{\mu}\boldsymbol{\mathbb{I}}\label{spinorialconnection}
\end{eqnarray}
needed to define
\begin{eqnarray}
&\boldsymbol{\nabla}_{\mu}\psi\!=\!\partial_{\mu}\psi
\!+\!\boldsymbol{\Omega}_{\mu}\psi\label{spincovder}
\end{eqnarray}
which is the spinorial covariant derivative.

Writing spinor fields in polar form does not only allow us to distill the spinor components into the real degrees of freedom, but it also provides the definition of the $\boldsymbol{S}$ matrix, which verifies
\begin{eqnarray}
&\boldsymbol{S}\partial_{\mu}\boldsymbol{S}^{-1}\!=\!i\partial_{\mu}\lambda\mathbb{I}
\!+\!\frac{1}{2}\partial_{\mu}\theta_{ij}\boldsymbol{\sigma}^{ij}\label{parameters}
\end{eqnarray}
where $\lambda$ is a generic complex phase and $\theta_{ij}\!=\!-\theta_{ji}$ are the six parameters of the Lorentz group. It is then possible to define
\begin{eqnarray}
&\partial_{\mu}\theta_{ij}\!-\!\Omega_{ij\mu}\!\equiv\!R_{ij\mu}\label{R}\\
&\partial_{\mu}\lambda\!-\!qA_{\mu}\!\equiv\!P_{\mu}\label{P}
\end{eqnarray}
which can be proven to be real tensors. The spin connection $\!\Omega_{ij\mu}\!$ carries information about gravity and coordinate systems while the derivative $\partial_{\mu}\theta_{ij}\!$ carries information about the coordinate system, and therefore $R_{ij\mu}$ carries information about gravity and coordinate systems. However, while independently non-tensorial quantities, their combination makes the non-tensorial spurious terms cancel, and the result is that $\!R_{ij\mu}$ is a real tensor. This is the reason why it is called \emph{tensorial connection}. Similarly, $qA_{\mu}$ contains information about electrodynamics and gauge phases while $\partial_{\mu}\lambda$ about gauge phases. While independently they are not gauge invariant, their combination $P_{\mu}$ is a real gauge-invariant vector. This is why it is called \emph{gauge-invariant vector momentum}. Due to their analogy, we will collectively call them tensorial connections, for simplicity \cite{Fabbri:2017pwp}. One can show that
\begin{eqnarray}
&\!\boldsymbol{\nabla}_{\mu}\psi\!=\!(\nabla_{\mu}\ln{\phi}\mathbb{I}
\!-\!\frac{i}{2}\nabla_{\mu}\beta\boldsymbol{\pi}
\!-\!iP_{\mu}\mathbb{I}\!-\!\frac{1}{2}R_{ij\mu}\boldsymbol{\sigma}^{ij})\psi
\label{decspinder}
\end{eqnarray}
from which
\begin{eqnarray}
&\nabla_{\mu}s_{i}\!=\!R_{ji\mu}s^{j}\label{ds}\\
&\nabla_{\mu}u_{i}\!=\!R_{ji\mu}u^{j}\label{du}
\end{eqnarray}
which are valid as general geometric identities. 
\subsection{Dynamical Equations}
The commutator of spinorial covariant derivatives can be used to define
\begin{eqnarray}
&R^{i}_{\phantom{i}j\mu\nu}\!=\!\partial_{\mu}\Omega^{i}_{\phantom{i}j\nu}
\!-\!\partial_{\nu}\Omega^{i}_{\phantom{i}j\mu}
\!+\!\Omega^{i}_{\phantom{i}k\mu}\Omega^{k}_{\phantom{k}j\nu}
\!-\!\Omega^{i}_{\phantom{i}k\nu}\Omega^{k}_{\phantom{k}j\mu}\\
&F_{\mu\nu}\!=\!\partial_{\mu}A_{\nu}\!-\!\partial_{\nu}A_{\mu}
\end{eqnarray}
which are the space-time and gauge curvatures.

It is straightforward to prove that
\begin{eqnarray}
&\!\!\!\!\!\!\!\!R^{i}_{\phantom{i}j\mu\nu}\!=\!-(\nabla_{\mu}R^{i}_{\phantom{i}j\nu}
\!-\!\!\nabla_{\nu}R^{i}_{\phantom{i}j\mu}
\!\!+\!R^{i}_{\phantom{i}k\mu}R^{k}_{\phantom{k}j\nu}
\!-\!R^{i}_{\phantom{i}k\nu}R^{k}_{\phantom{k}j\mu})\label{Riemann}\\
\!\!\!\!&qF_{\mu\nu}\!=\!-(\nabla_{\mu}P_{\nu}\!-\!\nabla_{\nu}P_{\mu})\label{Maxwell}
\end{eqnarray}
showing that the Riemann tensor can be written in terms of the tensorial connection while the Maxwell tensor can be written in terms of the gauge-invariant vector momentum. The tensorial connection and the gauge-invariant vector momentum are therefore the potentials of the gravitational and electrodynamic fields \cite{Fabbri:2018crr}. However, in absence of gravity or electrodynamics, when the curvatures vanish identically, differently from the connection and the gauge potential, which can always be vanished with a choice of frame or gauge, there is no way to vanish the tensorial connection and the gauge-invariant vector momentum, if they do not vanish identically already.

For the matter field, the dynamics is defined in terms of the Dirac spinor field equation
\begin{eqnarray}
&i\boldsymbol{\gamma}^{\mu}\boldsymbol{\nabla}_{\mu}\psi
\!+\!i\omega F_{\mu\nu}\boldsymbol{\sigma}^{\mu\nu}\psi\!-\!m\psi\!=\!0\label{D}
\end{eqnarray}
in which the $\omega$ term is an additional potential describing the coupling of the dipole moment of the spinor to an external field, which will be used to represent the potential of the harmonic oscillator later in this work.

It is now possible to substitute (\ref{decspinder}) into (\ref{D}) to write the Dirac spinor field equation in polar form. We then proceed to the Gordon decomposition by multiplying on the left with $\overline{\psi}$, $\overline{\psi}\boldsymbol{\gamma}^{a}$, $\overline{\psi}\boldsymbol{\sigma}^{ab}$, $\overline{\psi}\boldsymbol{\pi}$ and $\overline{\psi}\boldsymbol{\gamma}^a\boldsymbol{\pi}$ so to get $16$ equations, and then we split into real and imaginary parts getting $32$ real equations. Of these $32$ real equations, we must expect that $8$ taken together will be equivalent to the $8$ real components of the Dirac equation (\ref{D}). These $8$ real equations are those obtained by selecting the imaginary part of the contraction with $\boldsymbol{\gamma}^{a}$ and the real part of the contraction with $\boldsymbol{\gamma}^a\boldsymbol{\pi}$: multiplying the first by $\cos{\beta}$ and the second by $\sin{\beta}$ and adding them and multiplying the first by $\sin{\beta}$ and the second by $\cos{\beta}$ and substracting them produces the diagonalization that leads to
\begin{eqnarray}
\nonumber
&-2\omega F_{\mu\nu}u^{\nu}\sin{\beta}
\!-\!\omega \varepsilon_{\mu\rho\eta\sigma}F^{\rho\eta}u^{\sigma}\cos{\beta}+\\
\nonumber
&+\frac{1}{2}\varepsilon_{\mu\alpha\nu\iota}R^{\alpha\nu\iota}
\!-\!2P^{\iota}u_{[\iota}s_{\mu]}+\\
&+\nabla_{\mu}\beta\!+\!2s_{\mu}m\cos{\beta}\!=\!0 \label{dep1}\\
\nonumber
&2\omega F_{\mu\nu}u^{\nu}\cos{\beta}
\!-\!\omega \varepsilon_{\mu\rho\eta\sigma}F^{\rho\eta}u^{\sigma}\sin{\beta}+\\
\nonumber
&+R_{\mu a}^{\phantom{\mu a}a}
\!-\!2P^{\rho}u^{\nu}s^{\alpha}\varepsilon_{\mu\rho\nu\alpha}+\\
&+2s_{\mu}m\sin{\beta}\!+\!\nabla_{\mu}\ln{\phi^{2}}\!=\!0\label{dep2} 
\end{eqnarray}
which can be proven, in return, to derive the polar form of the Dirac spinor field equation. This proves the equivalence between (\ref{dep1}, \ref{dep2}) and (\ref{D}) itself. So the $4$ spinorial field equations, which are $8$ real field equations, can be converted into one vector field equation and one axial-vector field equation, specifying the first-order derivatives of the module and of the Yvon-Takabayashi angle, determining the dynamics of the real degrees of freedom \cite{Fabbri:2016laz}.
\section{Application To Two Systems}
The theory developed so far is general, but applications can also be studied so as to better understand what are the properties of the tensorial connections: our goal is to see what happens in the sourceless case, that is in situations where the energy density is not large enough to be a source of gravitation. We can assume that there is no gravity, a flat space-time, and an identically vanishing Riemann tensor (\ref{Riemann}). The tensorial connection can however still be different from zero. In this case we would have some non-trivial potential with no strength.

To prove that such a non-vanishing tensorial connection can have an effect on a relativistic quantum matter distribution, we consider explicit examples. To make our examples stronger, we will choose exact solutions of integrable potentials: one is given by the Coulomb potential, leading to the description of the hydrogen atom; and the other is given by the elastic potential, leading to the description of the harmonic oscillator.

Both cases are interesting because they account for all integrable potentials known in physics. In the following we start by reviewing the case of the hydrogen atom as it was treated in \cite{Fabbri:2018crr}. Then we consider the harmonic oscillator in the $3$-dimensional case as presented in \cite{mm-yfs}. 

The harmonic oscillator has not yet been studied in the polar form, and thus we will present it with more details. 
\subsection{Non-Trivial Integrable Cases}
\subsubsection{The Hydrogen Atom Model}
The case of the hydrogen atom is very widely known and can be found in common textbooks. 

The interaction is given in terms of the Coulomb potential, that is the temporal component of the gauge potential vector 
\begin{eqnarray}
&qA_{t}\!=\!-\alpha/r
\end{eqnarray}
where $\alpha\!=\!q^{2}$ is the fine-structure constant given in units in which it is the square of the electric charge.

Looking for solutions in stationary form $i\partial_{t}\psi\!=\!E\psi$ and with the choice of spherical coordinates 
\begin{eqnarray}
&\vec{r}\!=\!\left(\begin{array}{c}
\!r\sin{\theta}\cos{\varphi}\!\\
\!r\sin{\theta}\sin{\varphi}\!\\
\!r\cos{\theta}\!
\end{array}\right)
\end{eqnarray}
the Dirac spinor equations are written according to
\begin{eqnarray}
\nonumber
&(E\!+\!\frac{\alpha}{r})\!\left(\begin{array}{cc}
\!\mathbb{I} & \ \ 0 \\
\!0 & -\mathbb{I}\!
\end{array}\right)\!\psi
\!+\!\frac{i}{r}\!\left(\begin{array}{cc}
\!0 & \ \vec{\boldsymbol{\sigma}}\!\cdot\!\vec{r}\ \\
\!-\vec{\boldsymbol{\sigma}}\!\cdot\!\vec{r}\ & 0\!
\end{array}\right)\!\partial_{r}\psi-\\
&-\frac{i}{r^{2}}\!\left(\begin{array}{cc}
\!0 & \ \vec{\boldsymbol{\sigma}}\!\cdot\!\vec{r}\ \vec{\boldsymbol{\sigma}}\!\cdot\!\vec{L}\ \\
\!-\vec{\boldsymbol{\sigma}}\!\cdot\!\vec{r}\ \vec{\boldsymbol{\sigma}}\!\cdot\!\vec{L}\ & 0\!
\end{array}\right)\!\psi\!-\!m\psi\!=\!0\label{dee}
\end{eqnarray}
where
\begin{eqnarray}
&\vec{L}F\!=\!\left(\begin{array}{c}
\!i\sin{\varphi}\partial_{\theta}F\!+\!i\cot{\theta}\cos{\varphi}\partial_{\varphi}F\!\\
\!-i\cos{\varphi}\partial_{\theta}F\!+\!i\cot{\theta}\sin{\varphi}\partial_{\varphi}F\!\\
\!-i\partial_{\varphi}F\!
\end{array}\right)\label{ang}
\end{eqnarray}
for any function $F$, given in terms of the elevation and azimuthal angles. This form is well suited to study all cases where a separation of variables is possible.

We will focus on the ground-state, the $1S$ orbital.

In this case, defining the constant $\Gamma\!=\!\sqrt{1-\alpha^{2}}$ as well as the function 
$\Delta(\theta)\!=\!1/\!\sqrt{1\!-\!\alpha^{2}|\!\sin{\theta}|^{2}}$ of the elevation angle alone, it is possible to see that the energy is given by $E\!=\!m\Gamma$ and the spinor
\begin{eqnarray}
&\psi\!=\!\frac{1}{\sqrt{1+\Gamma}}r^{\Gamma-1}e^{-\alpha m r}e^{-iEt}\left(\begin{array}{c}
\!1\!+\!\Gamma\!\\
\!0\!\\
\!i\alpha\cos{\theta}\!\\
\!i\alpha\sin{\theta}e^{i\varphi}\!
\end{array}\right)\label{solution1}
\end{eqnarray}
is an exact solution of (\ref{dee}) with (\ref{ang}). To see this, one can insert (\ref{solution1}) into and (\ref{ang}) and (\ref{dee}) and check directly.

This is the standard treatment, but equations (\ref{dee}, \ref{ang}) are just the Dirac spinor equations (\ref{D}) for $\omega\!=\!0$ written in spherical coordinates
\begin{eqnarray}
&g_{tt}\!=\!1\\
&g_{rr}\!=\!-1\\
&g_{\theta\theta}\!=\!-r^{2}\\
&g_{\varphi\varphi}\!=\!-r^{2}|\!\sin{\theta}|^{2}
\end{eqnarray}
with connection
\begin{eqnarray}
&\Lambda^{\theta}_{\theta r}\!=\!\frac{1}{r}\\
&\Lambda^{r}_{\theta\theta}\!=\!-r\\
&\Lambda^{\varphi}_{\varphi r}\!=\!\frac{1}{r}\\
&\Lambda^{r}_{\varphi\varphi}\!=\!-r|\!\sin{\theta}|^{2}\\
&\Lambda^{\varphi}_{\varphi\theta}\!=\!\cot{\theta}\\
&\Lambda^{\theta}_{\varphi\varphi}\!=\!-\cot{\theta}|\!\sin{\theta}|^{2}
\end{eqnarray}
in the case in which the tetrad vectors are chosen to be
\begin{eqnarray}
&\!\!\!\!e^{0}_{t}\!=\!1\\
&\!\!\!\!e^{1}_{r}\!=\!\sin{\theta}\cos{\varphi}\ \ \ \ 
e^{2}_{r}\!=\!\sin{\theta}\sin{\varphi}\ \ \ \ 
e^{3}_{r}\!=\!\cos{\theta}\\
&\!\!\!\!e^{1}_{\theta}\!=\!r\cos{\theta}\cos{\varphi}\ \ \ 
e^{2}_{\theta}\!=\!r\cos{\theta}\sin{\varphi}\ \ \
e^{3}_{\theta}\!=\!-r\sin{\theta}\\
&\!\!\!\!e^{1}_{\varphi}\!=\!-r\sin{\theta}\sin{\varphi}\ \ \ \ 
e^{2}_{\varphi}\!=\!r\sin{\theta}\cos{\varphi}
\end{eqnarray}
and
\begin{eqnarray}
&\!\!\!\!e_{0}^{t}\!=\!1\\
&\!\!\!\!e_{1}^{r}\!=\!\sin{\theta}\cos{\varphi}\ \ \ \ 
e_{2}^{r}\!=\!\sin{\theta}\sin{\varphi}\ \ \ \ 
e_{3}^{r}\!=\!\cos{\theta}\\
&\!\!\!\!e_{1}^{\theta}\!=\!\frac{1}{r}\cos{\theta}\cos{\varphi}\ \ \ 
e_{2}^{\theta}\!=\!\frac{1}{r}\cos{\theta}\sin{\varphi}\ \ \
e_{3}^{\theta}\!=\!-\frac{1}{r}\sin{\theta}\\
&\!\!\!\!e_{1}^{\varphi}\!=\!-\frac{1}{r\sin{\theta}}\sin{\varphi}\ \ \ \ 
e_{2}^{\varphi}\!=\!\frac{1}{r\sin{\theta}}\cos{\varphi}
\end{eqnarray}
as the choice for which the spin connection vanishes.

Nevertheless, another specific choice is possible. It consists in taking the tetrad vectors as
\begin{eqnarray}
&\!\!\!\!e^{0}_{t}\!=\!\Delta\ \ \ \ 
e^{2}_{t}\!=\!-\alpha\sin{\theta}\Delta\\
&\!\!\!\!e^{1}_{r}\!=\!\Gamma\sin{\theta}\Delta\ \ \ \ 
e^{3}_{r}\!=\!\cos{\theta}\Delta\\
&\!\!\!\!e^{1}_{\theta}\!=\!r\cos{\theta}\Delta\ \ \ \ 
e^{3}_{\theta}\!=\!-\Gamma r\sin{\theta}\Delta\\
&\!\!\!\!e^{0}_{\varphi}\!=\!-\alpha r|\!\sin{\theta}|^{2}\Delta\ \ \ \ 
e^{2}_{\varphi}\!=\!r\sin{\theta}\Delta
\end{eqnarray}
and
\begin{eqnarray}
&\!\!\!\!e_{0}^{t}\!=\!\Delta\ \ \ \ 
e_{2}^{t}\!=\!\alpha\sin{\theta}\Delta\\
&\!\!\!\!e_{1}^{r}\!=\!\Gamma\sin{\theta}\Delta\ \ \ \ 
e_{3}^{r}\!=\!\cos{\theta}\Delta\\
&\!\!\!\!e_{1}^{\theta}\!=\!\frac{1}{r}\cos{\theta}\Delta\ \ \ \ 
e_{3}^{\theta}\!=\!-\frac{\Gamma}{r}\sin{\theta}\Delta\\
&\!\!\!\!e_{0}^{\varphi}\!=\!\frac{\alpha}{r}\Delta\ \ \ \ 
e_{2}^{\varphi}\!=\!\frac{1}{r\sin{\theta}}\Delta
\end{eqnarray}
which means that we are in the system of reference where the spinor field is in polar form.

We have then that
\begin{eqnarray}
&\beta\!=\!-\arctan{(\frac{\alpha}{\Gamma}\cos{\theta})}\label{b1}
\end{eqnarray}
and
\begin{eqnarray}
&\phi\!=\!r^{\Gamma-1}e^{-\alpha m r}/\sqrt{\Delta}\label{m1}
\end{eqnarray}
for the Yvon-Takabayashi angle and module.

Then we can compute
\begin{eqnarray}
&R_{t\varphi\theta}\!=\!-\alpha r \sin{\theta}\cos{\theta}|\Delta|^{2}\\
&R_{r\theta\theta}\!=\!-r(1\!-\!\Gamma|\Delta|^{2})\\
&R_{r\varphi\varphi}\!=\!-r|\!\sin{\theta}|^{2}\\
&R_{\theta\varphi\varphi}\!=\!-r^{2}\sin{\theta}\cos{\theta}
\end{eqnarray}
and
\begin{eqnarray}
&P_{t}\!=\!E\!+\!\alpha/r\\
&P_{\varphi}\!=\!-1/2
\end{eqnarray}
as it is well known for the momentum.

One can check that the pair of equations (\ref{dep1}, \ref{dep2}) is satisfied, as expected since (\ref{D}) is equivalent to (\ref{dep1}, \ref{dep2}).

For more details on the hydrogen atom we refer to \cite{Fabbri:2018crr}.

\subsubsection{The Harmonic Oscillator Model}
The case of the harmonic oscillator is also well known although its relativistic treatment is not so thoroughly investigated. In the following we will refer to \cite{mm-yfs}.

The interactions are given in terms of a coupling between the dipole moment of the spinor and an external field, like the one given in (\ref{D}):\begin{eqnarray}
&F_{\mu\nu}\!=\!v_{\mu}x_{\nu}\!-\!v_{\nu}x_{\mu}
\end{eqnarray}
with $v_{\mu}$ a time-like vector and $x_{\mu}$ the position vector. In the case we intend to study, the time-like vector will be chosen in the configuration in which only its temporal component remains and is normalized to unity.

We still look for solutions in the stationary form and in spherical coordinates, where (\ref{D}) is given by
\begin{eqnarray}
\nonumber
&E\!\left(\begin{array}{cc}
\!\mathbb{I} & \ \ 0 \\
\!0 & -\mathbb{I}\!
\end{array}\right)\!\psi
\!+\!\frac{i}{r}\!\left(\begin{array}{cc}
\!0 & \ \vec{\boldsymbol{\sigma}}\!\cdot\!\vec{r}\ \\
\!-\vec{\boldsymbol{\sigma}}\!\cdot\!\vec{r}\ & 0\!
\end{array}\right)\!\partial_{r}\psi-\\
\nonumber
&-\frac{i}{r^{2}}\!\left(\begin{array}{cc}
\!0 & \ \vec{\boldsymbol{\sigma}}\!\cdot\!\vec{r}\ \vec{\boldsymbol{\sigma}}\!\cdot\!\vec{L}\ \\
\!-\vec{\boldsymbol{\sigma}}\!\cdot\!\vec{r}\ \vec{\boldsymbol{\sigma}}\!\cdot\!\vec{L}\ & 0\!
\end{array}\right)\!\psi-\\
&-i\omega\!\left(\begin{array}{cc}
\! 0 & \vec{\boldsymbol{\sigma}}\!\cdot\!\vec{r}\\
\! \vec{\boldsymbol{\sigma}}\!\cdot\!\vec{r} & 0\!
\end{array}\right)\!\psi\!-\!m\psi\!=\!0\label{deo}
\end{eqnarray}
and as it is easy to see, this form is well suited for a separation of variables. However, we shall not implement this separation because it is known that this property does not hold for the harmonic oscillator, in the general case, when no non-relativistic limit is taken.

As before, we focus only on the ground-state. 

Defining the constant $a\!=\!(E\!-\!m)/2\omega$ together with the function 
$A(r,\theta)\!=\!\sqrt{r^{4}\!+\!a^{4}\!+\!2r^{2}a^{2}\cos{(2\theta)}}$ of the radial coordinate and elevation angle, one can see that the energy is given by $E^{2}\!=\!m^{2}\!+\!6\omega$ with the spinor given by
\begin{eqnarray}
&\psi\!=\!Ke^{-\frac{1}{2}\omega r^{2}}e^{-iEt}\left(\begin{array}{c}
\!r\cos{\theta}\!\\
\!r\sin{\theta}e^{i\varphi}\!\\
\!-ia\!\\
\!0\!
\end{array}\right)\label{solution2}
\end{eqnarray}
as an exact solution of (\ref{deo}) for any constant $K$.

Equations (\ref{deo}) are the Dirac spinor equations (\ref{D}) with no electric charge and written in spherical coordinates in the case in which the tetrad vectors are chosen as before.

And as before, another possibility is to Lorentz transform everything so to get the polar form. To this purpose, one first needs to implement a rotation along the third axis so as to perform a shift of $\varphi/2$ giving
\begin{eqnarray}
&\psi\!=\!Ke^{-\frac{1}{2}\omega r^{2}}e^{-i(Et-\frac{\varphi}{2})}\left(\begin{array}{c}
\!r\cos{\theta}\!\\
\!r\sin{\theta}\!\\
\!-ia\!\\
\!0\!
\end{array}\right)
\end{eqnarray}
in standard representation. With this solution we can calculate all bi-linear spinor quantities
\begin{eqnarray}
&\Theta\!=\!K^2 e^{-\omega r^{2}}(r^2-a^2)\\
&\Phi\!=\!K^2 e^{-\omega r^{2}} (2ar \cos{\theta})\\
&U^{0}\!=\!K^2 e^{-\omega r^{2}}(r^2 -a^2)\\
&U^{2}\!= \!K^2 e^{-\omega r^{2}} (2ar \sin{\theta}) \\
&S^{1}\!=\!K^2 e^{-\omega r^{2}} (2r^2 \sin{(2\theta)}) \\
&S^{3}\!=\!K^2 e^{-\omega r^{2}}(r^2 \cos{(2\theta)}+a^2)
\end{eqnarray}
and with $U^{1}\!=\!U^{3}\!=\!S^{0}\!=\!S^{2}\!=\!0$ identically. In order to force $U^{2}\!=\!S^{1}\!=\!0$ too, the only transformations of interest remain the boost along the second axis and the rotation around the second axis, given by
\begin{eqnarray}
\boldsymbol{B}_2\!=\!\begin{pmatrix}
\cosh{\xi} & 0 & \sinh{\xi} & 0 \\
0 & 1 & 0 & 0 \\
\sinh{\xi} & 0 & \cosh{\xi} & 0 \\
0 & 0 & 0 & 1
\end{pmatrix}
\end{eqnarray}
and
\begin{eqnarray}
\boldsymbol{R}_2\!=\!\begin{pmatrix}
1& 0 & 0 & 0 \\
0 & \cos{\chi} & 0 & \sin{\chi}\\
0 & 0 & 1 & 0 \\
0 & - \sin{\chi} & 0 & \cos{\chi} \
\end{pmatrix}
\end{eqnarray}
in terms of the rapidity
\begin{eqnarray}
\tanh{\xi}\!=\!\bigg( \frac{-2ar \sin \theta}{r^2 +a^2} \bigg)
\end{eqnarray}
and the angle
\begin{eqnarray}
\tan{\chi}\!=\!\bigg( \frac{-r^2 \sin( 2 \theta)}{r^2 \cos (2 \theta) +a^2} \bigg)
\end{eqnarray}
precisely because these are the rapidity and angle in terms of which $\boldsymbol{B}_2$ and $\boldsymbol{R}_2$ vanish $U^{2}$ and $S^{1}$ identically, respectively. This would mean that we have boosted into the rest frame and rotated the spin along the third axis, and therefore that we have written the spinor in polar form, which reads
\begin{eqnarray}
&\!\psi\!=\!\phi e^{-\frac{i}{2}\beta\boldsymbol{\pi}}
\boldsymbol{S}\left(\!\begin{tabular}{c}
$\sqrt{2}$\\
$0$\\
$0$\\
$0$
\end{tabular}\!\right)
\label{spinorST}
\end{eqnarray}
in standard representation. Here $\boldsymbol{S}\!=\!\boldsymbol{B}_2^{-1}\boldsymbol{R}_2^{-1}\boldsymbol{R}_3^{-1}$ with
\begin{eqnarray}
&\beta\!=\!\arctan{(\frac{2ar\cos{\theta}}{r^{2}-a^{2}})}\label{b2}
\end{eqnarray}
and
\begin{eqnarray}
&\phi\!=\!Ke^{-\frac{1}{2}\omega r^{2}}\sqrt{A/2}\label{m2}
\end{eqnarray}
for the Yvon-Takabayashi angle and module.

The same rapidity and angle, but for the real representation of Lorentz transformations, would boost and rotate tetrads so as to write them according to
\begin{eqnarray}
&\!\!\!\!\!\!\!\!e^{0}_{t}\!=\!(r^{2}\!+\!a^{2})A^{-1}\ \ \ \ 
e^{2}_{t}\!=\!-2ar\sin{\theta}A^{-1}\\
&\!\!\!\!e^{1}_{r}\!=\!-\sin{\theta}(r^{2}\!-\!a^{2})A^{-1}\ \ \ \ 
e^{3}_{r}\!=\!\cos{\theta}(r^{2}\!+\!a^{2})A^{-1}\\
&\!\!\!\!\!\!\!\!e^{1}_{\theta}\!=\!r\cos{\theta}(r^{2}\!+\!a^{2})A^{-1}\ \ \ \ 
e^{3}_{\theta}\!=\!r\sin{\theta}(r^{2}\!-\!a^{2})A^{-1}\\
&\!\!\!\!e^{0}_{\varphi}\!=\!-2ar^{2}|\!\sin{\theta}|^{2}A^{-1}\ \ \ \ 
e^{2}_{\varphi}\!=\!r\sin{\theta}(r^{2}\!+\!a^{2})A^{-1}
\end{eqnarray}
and
\begin{eqnarray}
&\!\!\!\!\!\!\!\!e_{0}^{t}\!=\!(r^{2}\!+\!a^{2})A^{-1}\ \ \ \ 
e_{2}^{t}\!=\!2ar\sin{\theta}A^{-1}\\
&\!\!\!\!e_{1}^{r}\!=\!-\sin{\theta}(r^{2}\!-\!a^{2})A^{-1}\ \ \ \ 
e_{3}^{r}\!=\!\cos{\theta}(r^{2}\!+\!a^{2})A^{-1}\\
&\!\!\!\!\!\!\!\!e_{1}^{\theta}\!=\!\frac{1}{r}\cos{\theta}(r^{2}\!+\!a^{2})A^{-1}\ \ \ \ 
e_{3}^{\theta}\!=\!\frac{1}{r}\sin{\theta}(r^{2}\!-\!a^{2})A^{-1}\\
&\!\!\!\!e_{0}^{\varphi}\!=\!2aA^{-1}\ \ \ \ 
e_{2}^{\varphi}\!=\!\frac{1}{r\sin{\theta}}(r^{2}\!+\!a^{2})A^{-1}
\end{eqnarray}
and in terms of which it is now possible to calculate $R_{i j \mu}$ with (\ref{R}) getting
\begin{eqnarray}
&R_{t\varphi\theta}\!=\!-2ar^{2}\sin{\theta}\cos{\theta}(r^{2}\!+\!a^{2})A^{-2}\\
&R_{r\theta\theta}\!=\!-2r^{3}[r^{2}\!+\!a^{2}\cos{(2\theta)}]A^{-2}\\
&R_{t\varphi r}\!=\!2ar|\!\sin{\theta}|^{2}(r^{2}\!-\!a^{2})A^{-2}\\
&R_{r\theta r}\!=\!-2a^{2}r^{2}\sin{(2\theta)}A^{-2}\\
&R_{r\varphi\varphi}\!=\!-r|\!\sin{\theta}|^{2}\\
&R_{\theta\varphi\varphi}\!=\!-r^{2}\sin{\theta}\cos{\theta}
\end{eqnarray}
while we also have
\begin{eqnarray}
&P_{t}\!=\!E\\
&P_{\varphi}\!=\!-1/2
\end{eqnarray}
as it is again well known for the momentum.

One can see that the pair of equations (\ref{dep1}, \ref{dep2}) is satisfied, as expected since (\ref{D}) is equivalent to (\ref{dep1}, \ref{dep2}).

With the case of the harmonic oscillator completed it is now possible to compare the two physical examples.
\subsection{The Comparison In Parallel}
\subsubsection{Bi-Linear Invariant Quantities}
In order to make the comparison meaningful, it is easier to consider quantities that are free of any superfluous information. For this reason, we focus on scalars, since they are the only quantities that can be invariant while still being non-trivial. To make the comparison easy to read, in the following, we express the considered quantities for the hydrogen atom first and for the harmonic oscillator just below.

To begin, the Yvon-Takabayashi angles are
\begin{eqnarray}
&\beta\!=\!\arctan{(-\frac{\alpha}{\Gamma}\cos{\theta})}\\
&\beta\!=\!\arctan{(\frac{2ar\cos{\theta}}{r^{2}-a^{2}})}
\end{eqnarray}
and the modules are
\begin{eqnarray}
&\phi\!=\!r^{\Gamma-1}e^{-\alpha m r}/\sqrt{\Delta}\\
&\phi\!=\!Ke^{-\frac{1}{2}\omega r^{2}}\sqrt{A/2}
\end{eqnarray}
where some information already becomes visible: for instance, the Yvon-Takabayashi angle must be an odd function of $\cos{\theta}$ because of its pseudo-scalar character, and we see no radial dependence in the Yvon-Takabayashi angle in concomitance with the separability of variables of the module in the case of the hydrogen atom, while no such feature exists for the harmonic oscillator.

This is obvious from the fact that whenever the separability of variables is demanded, the module must be a product of the form $\phi\!=\!R(r)Y(\theta)$ while at the same time the Yvon-Takabayashi angle must be a sum of the form $\beta\!=\!S(r)\!+\!Z(\theta)$ since it is the argument of an exponential function. Because under parity the Yvon-Takabayashi angle flips its sign, we then must have $S\!=\!0$ necessarily.

It should however be noticed that when the separation of variable does not hold, as for the harmonic oscillator, the radial dependence can carry surprises: for instance, it is easy to see that at $r\!=\!a$ the Yvon-Takabayashi angle is equal to $\pm\pi/2$. This defines the boundary between the regions where $\cos{\beta}$ is positive and regions where it is negative. Because of this, the sphere of radius $a$ is the limit through which the scalar density $\Phi$ changes sign.

The five scalars coming from the squares of the tensorial connections are given by
\begin{eqnarray}
&\!\!R_{ac}^{\phantom{ac}c}R^{ai}_{\phantom{ai}i}\!=\!-\frac{1}{r^{2}}\!
\left[(2\!-\!\Gamma\Delta^{2})^{2}\!+\!|\!\cot{\theta}|^{2}\right]\\
&\!\!R_{ac}^{\phantom{ac}c}R^{ai}_{\phantom{ai}i}\!=\!4(a^{2}\!-\!2r^{2})A^{-2}
\!-\!\left|\frac{1}{r\sin{\theta}}\right|^{2}\\
\nonumber\\
&\!\!\!\!\frac{1}{4}R_{ijk}R^{abc}\varepsilon^{pijk}\varepsilon_{pabc}
\!=\!-\frac{1}{r^{2}}\alpha^{2}|\!\cos{\theta}|^{2}\Delta^{4}\\
&\!\!\!\!\frac{1}{4}R_{ijk}R^{abc}\varepsilon^{pijk}\varepsilon_{pabc}\!=\!-4a^{2}A^{-2}\\
\nonumber\\
&\!\!\!\!\!\!\!\!\frac{1}{2}R_{ijk}R^{ijk}\!=\!\frac{1}{r^{2}}\!
\left[\alpha^{2}|\!\cos{\theta}|^{2}\Delta^{4}\!-\!(1\!-\!\Gamma\Delta^{2})^{2}
\!-\!\frac{1}{|\!\sin{\theta}|^{2}}\right]\\
&\!\!\!\!\!\!\!\!\frac{1}{2}R_{ijk}R^{ijk}\!=\!4(a^{2}\!-\!r^{2})A^{-2}
\!-\!\left|\frac{1}{r\sin{\theta}}\right|^{2}\\
\nonumber\\
&\!\!\frac{1}{2}R_{pq}^{\phantom{pq}q}R_{ijk}\varepsilon^{pijk}\!=\!\frac{1}{r^{2}}\alpha\cos{\theta}\Delta^{2}(2\!-\!\Gamma\Delta^{2})\\
&\!\!\frac{1}{2}R_{pq}^{\phantom{pq}q}R_{ijk}\varepsilon^{pijk}\!=\!8ar\cos{\theta}A^{-2}\\
\nonumber\\
&\!\!\!\!\frac{1}{4}R_{ijc}R_{pq}^{\phantom{pq}c}\varepsilon^{ijpq}
\!=\!\frac{2}{r^{2}}\alpha\cos{\theta}\Delta^{2}(1\!-\!\Gamma\Delta^{2})\\
&\!\!\!\!\frac{1}{4}R_{ijc}R_{pq}^{\phantom{pq}c}\varepsilon^{ijpq}
\!=\!8ar\cos{\theta}A^{-2}
\end{eqnarray}
and something interesting is also emerging here: while in the large-$r$ regime, in both cases, all scalars tend to zero, in the small-$r$ region, for the hydrogen atom all scalars behave as $1/r^{2}$ whereas for the harmonic oscillator only $R_{ac}^{\phantom{ac}c}R^{ai}_{\phantom{ai}i}$ and $R_{ijk}R^{ijk}$ behave as $1/r^{2}$. As it is expected, both pseudo-scalars tend to zero with a linear behaviour in the radial coordinate but $R_{ijk}R^{abc}\varepsilon^{pijk}\varepsilon_{pabc}\!\approx\!-16/a^{2}$ and the fact that some scalar tends to a non-vanishing constant looks a very astonishing circumstance.

This is a consequence of the fact that for the hydrogen atom all scalars must have the same radial behavior to ensure dimensional consistency while for the harmonic oscillator the constant $a$ has the dimension of a length and can therefore be substituted to the radial coordinate in some expressions. Nevertheless, for cases in which there is a natural constant with the dimension of a length, we do not think that it is possible to guess the actual radial behavior. There is in fact no {\it a priori} difference between the three scalars and still one of them has a radial behavior that is very different from the one of the others.

\subsubsection{Energy Density Tensor Components}
Albeit scalars are the invariants of the theory, it might also be instructive to see what happens for a non-scalar quantity. Even if we are considering situations where the energy is not large enough to be a relevant source for the gravitational field, it can still be different from zero and as such, it may contain some interesting information.

The energy density tensor which is the source term of the Einstein field equations is given, in polar form, according to
\begin{eqnarray}
&T_{qa}\!=\!2\phi^{2}(s_{q}\nabla_{a}\beta/2\!+\!u_{q}P_{a}
\!+\!\frac{1}{4}\varepsilon_{kijq}s^{k}R^{ij}_{\phantom{ij}a})
\end{eqnarray}
and it results in
\begin{eqnarray}
&T_{tt}\!=\!2\phi^{2}\Delta(E\!+\!\alpha/r)\\
&T_{tt}\!=\!2\phi^{2}A^{-1}(r^{2}\!+\!a^{2})E\\
\nonumber\\
&T_{t\varphi}\!=\!-\phi^{2}\Delta(1\!-\!\Gamma)|\!\sin{\theta}|^{2}\\
&T_{t\varphi}\!=\!-2\phi^{2}A^{-1}r^{2}|\!\sin{\theta}|^{2}\\
\nonumber\\
&T_{rr}\!=\!0\\
&T_{rr}\!=\!2\phi^{2}A^{-1}a\\
\nonumber\\
&T_{\theta\theta}\!=\!\phi^{2}\Delta\alpha r\\
&T_{\theta\theta}\!=\!2\phi^{2}A^{-1}ar^{2}\\
\nonumber\\
&T_{\varphi t}\!=\!-2\phi^{2}\Delta\alpha r|\!\sin{\theta}|^{2}(E\!+\!\alpha/r)\\
&T_{\varphi t}\!=\!-4\phi^{2}A^{-1}ar^{2}|\!\sin{\theta}|^{2}E\\
\nonumber\\
&T_{\varphi\varphi}\!=\!\phi^{2}\Delta\alpha r|\!\sin{\theta}|^{2}\\
&T_{\varphi\varphi}\!=\!2\phi^{2}A^{-1}ar^{2}|\!\sin{\theta}|^{2}
\end{eqnarray}
in which an obvious lack of symmetry can be noticed in the fact that in the case of the hydrogen atom there is no radial-radial component. This is once again a consequence of the separability of variables.

It is possible to compute the traces, which give
\begin{eqnarray}
&T\!=\!2\phi^{2}\Delta E\\
&T\!=\!2\phi^{2}A^{-1}(r^{2}E\!-\!a^{2}m)
\end{eqnarray}
and exhibit an interesting property: while for the hydrogen atom the large-$r$ and small-$r$ behaviors are the same, for the harmonic oscillator the large-$r$ behavior is $T\!=\!2\phi^{2}E$ but the small-$r$ behavior becomes $T\!=\!-2\phi^{2}m$ flipping the sign of the scalar trace of the energy density.

So the sphere of radius $a\sqrt{m/E}$ defines the limit through which the scalar trace of the energy density $T$ changes from positive values to negative values. Because the trace is such that $T\!=\!\mathscr{L}+m\overline{\psi}\psi$ with $\mathscr{L} $ the Lagrangian functional, we may think at the energy trace as what encodes information about the total energy.
\section{The Tensorial Connections}
In the first section we have seen that $R_{ijk}$ and $P_{a}$ have the character of connections while being true tensors: $R_{ijk}$ is the tensorial connection in a strict sense since it is directly related to the Lorentz transformation while $P_{a}$ is called the gauge-invariant vector momentum to highlight its relation to the gauge transformations. Although the concept of a tensorial connection seems a contradiction, because connections can be vanished with a choice of frame whereas tensors cannot, this should not appear as something drastically new: the orbital angular momentum can be vanished when calculated in specific points but the spin cannot. The tensorial connection and the spin share this property of being truly covariant.

However, tensorial connections still behave as connections in their lacking of couplings to sources. In fact, the components of $R_{ijk}$ and $P_{a}$ might well be different from zero but their curvatures are vanishing if, respectively, no gravity or no electrodynamic phenomenon is present.

Situations where some physical effects can be ascribed to potentials that are present (as non-zero connections) despite having no strength (since they have zero curvature) is something that may be strange for $R_{ijk}$ but for $P_{a}$ is what we already know as Aharonov-Bohm effect.

For $P_{a}$ the technicalities can be worked out by taking expression (\ref{P}) and integrating it as
\begin{eqnarray}
\int_{\gamma}(P_{\mu}\!+\!qA_{\mu})dx^{\mu}\!=\!
\int_{\gamma}\partial_{\mu}\lambda dx^{\mu}\!=\!
\int_{\gamma}d\lambda\!=\!
\Delta \lambda
\end{eqnarray}
along the trajectory $\gamma$, and where the last term is just the difference of phase between the starting and the ending points. Similarly, from (\ref{R}) we get
\begin{eqnarray}
\int_{\gamma}(R_{ij\mu}\!+\!\Omega_{ij\mu})dx^{\mu}\!=\!
\int_{\gamma}\partial_{\mu}\theta_{ij}dx^{\mu}\!=\!
\int_{\gamma}d\theta_{ij}\!=\!\Delta\theta_{ij}
\end{eqnarray}
in total analogy with the case above. Recall that Lorentz indices designate quantities that are tensor under a (local) Lorentz transformation but scalar under coordinate transformations. The integral is therefore well-defined.

When the spinor is in polar form, (\ref{ds}, \ref{du}) reduce to
\begin{eqnarray}
&\nabla_{\mu}s_{i}\!=\!R_{3i\mu}\\
&\nabla_{\mu}u_{i}\!=\!R_{0i\mu}
\end{eqnarray}
showing that the dynamics of the velocity vector or of the spin axial-vector is determined only by those components of $R_{ij\mu}$ for which the first index is equal to either zero or three. The antisymmetry in the first two indices implies that any of the first two indices has to be either zero or three, so that $R_{12\mu}$ never appears. This makes this component somehow analogous to the momentum since $P_{\mu}$ never appears in the dynamics of the velocity vector and of the spin axial-vector in the first place.

This is in line with the fact that for a spinor that is an eigenstate of the spin, that is for rotations around the third axis, as the one in the polar form, rotations around the third axis have the same effect of gauge shifts: in fact, a suitable rotation around the third axis generates a component of $R_{12\mu}$ which is related by $R_{12\mu}\!\equiv\!-2P_{\mu}$ to the momentum generated by an equivalent gauge shift. 

If the trajectory is a close circuit, $\Delta\theta$ is just a whole turn times an integer
\begin{eqnarray}
\oint_{\gamma}(P_{\mu}\!+\!qA_{\mu})dx^{\mu}\!=\!2\pi n
\end{eqnarray}
and analogously for $\Delta\theta_{12}$ we have
\begin{eqnarray}
\oint_{\gamma}(R_{12\mu}\!+\!\Omega_{12\mu})dx^{\mu}\!=\!-4\pi n
\end{eqnarray}
where $n$ is usually called winding number.
\subsection{Discretizing The Connection}
If we consider the free cases, requiring the electromagnetic field to vanish means that
\begin{eqnarray}
\oint_{\gamma}P_{\mu}dx^{\mu}\!=\!2\pi n
\end{eqnarray}
while requiring the gravitational field to vanish means that it is always possible to find a frame where
\begin{eqnarray}
\oint_{\gamma}R_{12\mu}dx^{\mu}\!=\!-4\pi n
\end{eqnarray}
as it is clear because of the formal analogy. Whereas the former is clearly the Born rule for discretizing momenta in closed orbits, the latter should be regarded as the Born rule for discretizing some components of the connection in closed orbits. Such an occurrence brings about an important point in the discussion around the quantization of gravitational degrees of freedom, because the tensorial connection is precisely where the geometrical information is encoded. The process of discretization is entirely independent on the structure of the tensorial connection.

Much in the same way in which tensorial connections can be discrete in the free case, the same might happen even if gravity were present. In this case the quantization would happen on the gravitational degrees of freedom. We do not claim that this approach solves the long standing problem of quantum gravity. It might however give some hints about the fundamentally quantum nature of some geometrical degrees of freedom.
\subsection{Aharonov-Bohm Effects}
If the gauge-invariant vector momentum and the tensorial connection happen to vanish, and we choose a close circuit to be the boundary of a given surface $\gamma\!=\!\partial S$, then
\begin{eqnarray}
q\oint_{\partial S}\vec{A}\!\cdot\!d\vec{x}\!=\!2\pi n
\end{eqnarray}
and analogously
\begin{eqnarray}
\oint_{\partial S}\vec{\Omega}_{12}\!\cdot\!d\vec{x}\!=\!-4\pi n
\end{eqnarray}
in which we accounted for the spatial parts only. Using the Stokes theorem we obtain
\begin{eqnarray}
q\int\!\!\!\!\int_{S}\mathrm{rot}\vec{A}\!\cdot\!d\vec{S}\!=\!2\pi n \label{em}
\end{eqnarray}
and analogously
\begin{eqnarray}
\int\!\!\!\!\int_{S}\mathrm{rot}\vec{\Omega}_{12}\!\cdot\!d\vec{S}\!=\!-4\pi n \label{g}
\end{eqnarray}
where we now have fluxes on the left-hand side. While the former is recognized to be the condition giving rise to the Aharonov-Bohm effect, the latter should be interpreted as the condition giving rise to the gravitational analogous of the Aharonov-Bohm effect. This would not only entail the quantization of the electromagnetic as well as of the gravitational fluxes, as discussed above. But it also means that there can be a phase-shift in the wave function of the matter field due to the electromagnetic as well as to the gravitational potentials even in regions with neither electromagnetic nor gravitational forces.

In fact, writing (\ref{spinor}) in the form
\begin{eqnarray}
&\psi\!=\!\boldsymbol{S}\psi_{\mathrm{pol}}
\end{eqnarray}
where $\psi_{\mathrm{pol}}$ is the spinor in full polar form, we have that
\begin{eqnarray}
&\boldsymbol{S}\!=\!e^{-i\lambda}e^{-\frac{1}{2}\theta_{ij}\boldsymbol{\sigma}^{ij}}\label{S}
\end{eqnarray}
in terms of one phase-shift of abelian type in $\lambda$ and another of non-abelian type in $\theta_{12}$ which, according to the above (\ref{em}, \ref{g}), can be present even in regions where no electrodynamic or gravity are present. However, electrodynamics or gravity must be present in nearby regions so to let the fluxes be non-zero at least somewhere.

The analogy of the two types of Aharonov-Bohm effect, electrodynamic and gravitational, can be appreciated in its full extent in the fact that in (\ref{S}) both abelian gauge phase and third-axis rotation angle have identical impact on the structure of the spinor field matter distribution.

Nevertheless, it is important to stress that the usual electrodynamic Aharonov-Bohm effect parallels only one of the six vector potentials describing the gravitational Aharonov-Bohm effect, and therefore the latter is inevitably richer in potential physical applications.
\section{Special Approximations}
As concluding remarks, we would like to investigate what happens in the case of specific limits. A first approximation is the one for which the two coupling constants are small: in such a case, the above solution for the hydrogen atom automatically reduces to the non-relativistic solution for the considered system. Instead, the solution for the harmonic oscillator has $a\!\approx\!\frac{3}{2m}$ which reduces to the non-relativistic solution only in the case of large masses. In fact, even if the mass is large, it would still be possible to consider radial distances small enough, and the non-relativistic approximation still fails.

In fact, quite generally, for the harmonic oscillator we can always find regions where relativistic effects cannot be suppressed. To see this, just consider the scalar quantity $\cos{\beta}$ and the energy trace $T$. The first changes sign on the sphere of radius $a$ and the second changes sign on the sphere of radius $a\sqrt{m/E}$ with $a\!>\!a\sqrt{m/E}$ since $\omega$ is positive. For small values of $\omega$, we can expand the energy and write it according to
\begin{eqnarray}
T\!\approx\!2\phi^{2}\left(m\cos{\beta}\!+\!\frac{3\omega r^{2}}{Am}\right)
\end{eqnarray}
which isolates the kinetic energy $m\cos{\beta}$ from the potential energy $3\omega r^{2}A^{-1}/m$. The kinetic energy becomes negative across the sphere of radius $a\sqrt{m/E}$ and it becomes negative and large enough so to overcome the positive potential and make the total energy negative as well across the sphere of radius $a\!>\!a\sqrt{m/E}$. Apart from this shift due to the potential, the reason for which both the energy and the module become negative is the same, that is the fact that $\cos{\beta}$ becomes negative. As $\cos{\beta}\!\rightarrow\!-1$ then $\beta\!\rightarrow\!\pi$ which means that left-handed and right-handed chiral parts are in maximal phase opposition with respect to one another. The deep interpretation of such unusual new effects is still to be understood but, at the heuristic level, calculation of observables are in principle possible.

Because the Yvon-Takabayashi angle is what describes the differences between the two chiral parts even in the rest frame, it can be interpreted as what describes the internal dynamics of spinor fields. Thus, close to the center of the matter distribution, where $\beta$ tends to its maximal value, there appears a region where the internal dynamics is dominant. This is the region where relativistic effects can never be suppressed, as we argued above.

Such an internal dynamics is confined within a sphere whose radius can be evaluated, for small values of $\omega$, to be approximately one fourth of the Compton wavelength.

From the viewpoint of ordinary QFT, this is a strange occurrence as the scalar density $\Phi$ is always assumed to be strictly positive in QFT. This implies that the harmonic oscillator has solutions which, as fields, cannot be quantized, or at least not with usual methods.

We will not deal, however, with second quantization.
\section{Conclusion}
In this work, we have shown that when the spinor fields are written in polar form, it becomes possible to define a pair of objects that contain the very same information of the space-time connection and the gauge potential but which are covariant under Lorentz and phase transformations: they are called tensorial connection and the gauge-invariant vector momentum. We have discussed that they are generally non-zero even when they have neither space-time curvature nor gauge curvature: this means that they can have effects even when sourceless. Although this may look surprising, we have shown that it consistently happens in specific cases, such as the Coulomb and elastic potentials. A final comparison between the hydrogen atom and the harmonic oscillator was also performed, in particular for the scalars and for the energy density tensor.

The fact that there could be non-trivial effects even when considering sourceless actions is not new, since a phase shift can occur in what is known as the Aharonov-Bohm effect. We have shown that such a phenomenon occurs not only for the gauge-invariant vector momentum but also for the tensorial connection. To highlight this, we have built a parallel between the two cases. We have also underlined that as the Aharonov-Bohm effect can entail information about the quantization of electromagnetic fluxes, the gravitational version of the Aharonov-Bohm effect may encode information about the quantization of at least some of gravitational fluxes.

We have concluded with comments on non-relativistic limits, and in particular we have underlined the fact that for the harmonic oscillator it is not possible to get non-relativistic approximations in regions that are too close to the center of the matter distribution because these are the regions where the internal dynamics is dominant.

The fact that for the harmonic oscillator, both the energy and the module become negative seems to lead to some conceptual problems in the perspective of QFT, since several results of QFT are based on assumptions that we show not to be fulfilled in general. For example, some hypotheses of spin-statistic theorems, like the positivity of energy and norms, should be questioned when harmonic oscillations are taken into account. Nevertheless, while critical in QFT, these features of the harmonic oscillator are a consequence of exact solutions in presence of elastic potentials within the Dirac equation, and so there does not seem to be much room for improvement.

The only possibility could be that the problems come from the elastic potential, but the elastic potential is just a dipole coupling to an external tensor field, like the one that occurs in presence of radiative processes.

We leave such considerations, and possible experimental signatures, for a future work.

\end{document}